\documentclass[12pt]{article}

\usepackage[margin=1in]{geometry}
\usepackage{amsmath}
\usepackage{amssymb}
\usepackage{graphicx}
\usepackage{booktabs}
\usepackage{natbib}
\usepackage{hyperref}
\usepackage{float}
\usepackage{setspace}
\usepackage{microtype}
\usepackage{caption}
\usepackage{subcaption}
\usepackage{xcolor}
\usepackage{soul}
\usepackage{array}
\usepackage{makecell,threeparttable,siunitx,adjustbox}
\usepackage{tikz}
\usepackage{listings}
\usetikzlibrary{positioning}
\hypersetup{
    colorlinks=true,
    linkcolor=blue,
    citecolor=blue,
    urlcolor=blue
}

\definecolor{sectioncolor}{RGB}{0,0,139}
\definecolor{placeholdercolor}{RGB}{139,0,0}
\definecolor{instructioncolor}{RGB}{0,100,0}

\renewcommand{\widehat}[1]{\hat{#1}}

\newcommand{\E}{\mathbb{E}}

\title{Leveraging LLMs to Improve Experimental Design: \\
A Generative Stratification Approach}
\author{George Gui\thanks{Columbia Business School. Email: zg2467@gsb.columbia.edu} \and Seungwoo Kim\thanks{Columbia Business School. Email: sk5623@gsb.columbia.edu}}
\date{September 29, 2025}

\begin{document}

\maketitle

\begin{abstract}
Pre-experiment stratification, or blocking, is a well-established technique for designing more efficient experiments and increasing the precision of the experimental estimates. However, when researchers have access to many covariates at the experiment design stage, they often face challenges in effectively selecting or weighting covariates when creating their strata. This paper proposes a \emph{Generative Stratification} procedure that leverages Large Language Models (LLMs) to synthesize high-dimensional covariate data to improve experimental design. We demonstrate the value of this approach by applying it to a set of experiments and find that our method would have reduced the variance of the treatment effect estimate by $10\%-50\%$ compared to simple randomization in our empirical applications. When combined with other standard stratification methods, it can be used to further improve the efficiency. Our results demonstrate that LLM-based simulation is a practical and easy-to-implement way to improve experimental design in covariate-rich settings. 
\end{abstract}

\newpage 

\section*{Introduction}
\addcontentsline{toc}{section}{Introduction}

Randomized controlled trials (RCTs) are the gold standard for causal inference across social and biomedical sciences, yet experiments can be costly \citep{duflo2007using} or require large sample sizes \citep{lewis2015unfavorable}. Pre-experiment stratification \citep{fisher1926arrangement}, also known as blocking, is a well-known way to improve statistical efficiency at the design stage \citep{bruhn2009pursuit, athey2017econometrics}. By grouping similar units in the same strata and randomizing assignment within the strata, the experiment is more balanced and hence more accurate.

Although the theoretical benefits of stratification are well-established \citep{athey2017econometrics} and its use has grown in social sciences (for example, among more than 10{,}000 RCTs in the AEA RCT Registry, more than 2{,}000 are stratified), further adoption has been hindered by several challenges. Stratification is sometimes infeasible because researchers do not observe enough informative covariates prior to the experiment. However, in the age of big data, it is increasingly common to observe many pre-treatment covariates that range from numeric variables to high-dimensional categorical variables and even unstructured variables such as text (e.g., interview transcripts, survey responses). This high-dimensional space presents distinct computational and methodological challenges for experiment design, given that it is unclear how to effectively select or weight different covariates. In some cases, researchers may lack the prior knowledge to choose which covariates are more important and how to transform high-dimensional covariates into a smaller set of informative signals on which to create useful strata. As a result, researchers either focus on one or two variables and leave the potential efficiency gains from other covariates on the table, or include all variables \citep{greevy2004optimal, morgan2012rerandomization} which performs poorly when many variables are irrelevant. In other cases, although researchers may understand how different variables jointly affect the outcome, it is challenging to evaluate each variable, examine interactions, and weight them appropriately, to the point where the mental cost is too high to justify its benefit.\footnote{For example, Bai (2022) documents that in several AEA-published papers, stratification as implemented by the authors provided essentially no efficiency gain.}

This paper proposes \emph{Generative Stratification} as a method to improve experimental design by synthesizing high-dimensional covariate information into an effective prognostic score using a large language model (LLM). The prognostic score is motivated by \cite{bai2022optimality}, who proves that when the treatment probability is one-half, the optimal stratification orders units by the sum of potential outcomes, $g(x)=\E[Y_i(1)+Y_i(0)\mid X_i=x]$, and forms strata along this ordering. A practical challenge is that neither $Y_i(0)$ nor $Y_i(1)$ is observed at the design stage; our core idea is to approximate this score by using an LLM to generate predicted potential outcomes for each experimental unit based on observable characteristics and the experimental context, and then to use these predictions to form strata that reduce variance.

We consider our approach an effective and safe application of LLM simulation that improves efficiency while preserving unbiasedness. The approach exploits several unique capabilities of modern LLMs that make them particularly suited for this task. First, LLMs can process heterogeneous data types simultaneously—demographic information, textual descriptions, behavioral histories, and contextual factors—without requiring extensive feature engineering or standardization. This flexibility allows researchers to incorporate all available pre-treatment information regardless of its format or structure, including high-dimensional categorical variables and unstructured text. Second, LLMs leverage vast pre-training corpora that encode complex relationships between individual characteristics and behaviors across many domains. This accumulated knowledge enables them to identify predictive patterns that might not be obvious ex ante or would require extensive domain expertise and human effort. Third, LLMs have the potential to incorporate contextual understanding of the experimental intervention itself, using natural language descriptions of treatment and control conditions to inform their predictions about how different types of units might respond.

The method has limited risk because effective stratification does not require perfect predictions; it requires a score that correlates with the (unknown) optimal score. Randomization within strata preserves unbiasedness, so even if LLM predictions are imperfect or biased, they do not induce bias in the treatment effect estimator. Thus, as long as the LLM-based score is correlated with the underlying true score, it improves experimental efficiency; when the correlation is weak, gains may be modest, but because this choice affects only the design stage, the final experiment remains valid due to randomization and the estimate remains unbiased. 

\section*{Literature Review}
\label{sec:background}
\addcontentsline{toc}{section}{Literature Review}

Our paper contributes to the literature of experimental design and stratification by proposing a practical and scalable way to implement stratification. The construction of our LLM-simulated prognostic score is motivated by \cite{bai2022optimality}, who demonstrates that for standard experiments with a 50/50 assignment, the optimal stratification design is obtained by ordering units on the score $g_i = Y_i(0) + Y_i(1)$ and then forming small strata (e.g., pairs).
A practical challenge is that neither $Y_i(0)$ nor $Y_i(1)$ is observed at the design stage. Our paper advances implementation by providing a practical and scalable way to incorporate high-dimensional or unstructured covariates into a single prognostic score. Rather than requiring researchers to hand-pick or weight covariates—which is impractical when many variables are available—we use an LLM to synthesize numeric, categorical, and textual features into a one-dimensional prognostic score. This directly addresses a core challenge emphasized in applied work: how to \emph{select and combine} multiple pre-treatment covariates into an effective stratification variable when no single baseline predictor dominates \citep{bruhn2009pursuit}. Our approach is complementary to existing, more theory-agnostic strategies that aggregate covariates through multivariate balance metrics or algorithmic rules—such as rerandomization on a prespecified balance criterion \citep{morgan2012rerandomization} and optimal multivariate matching (e.g., Mahalanobis) \citep{greevy2004optimal}. While our empirical applications focus on matched pairs, the LLM-generated score can be used to form strata of other sizes or serve as an additional covariate within existing schemes. Compared to other stratification designs that do not model how different covariates affect outcomes, our method complements them by incorporating plausible prior information in a scalable way, which is especially valuable when there are many moderate predictors but no obvious stratifying variable \citep[see also][]{athey2017econometrics}. In sum, we provide a stratification procedure grounded in econometric theory that bridges the gap between optimal-score results and their practical implementation.

Our paper contributes to the LLM simulation literature by demonstrating a safe, valuable application of LLM simulation in experimental design. While prior work shows LLMs have the potential to simulate human behavior \citep{aher_using_2023, argyle_out_2023, dillion_can_2023, horton_large_2023, brand_using_2023, park_generative_2023, arora2024express, goli2024frontiers, hewitt2024predicting, qin2024aiturk, binz2025foundation, toubia2025twin}, LLM simulations often exhibit biases \citep{goli_language_2023, santurkar2023whose, gui2023challenge, motoki2024more, brucks2025prompt} and generate unstable results \citep{mohammadi2024explaining, gao2025take}. These reliability concerns make practitioners hesitant to adopt such methods. We identify experimental stratification as an application where imperfect and biased predictions still yield efficiency gains without biasing final experimental estimates. Our paper relates to other recent work on using LLMs to safely improve efficiency. \cite{wang2024large} demonstrates the value of LLM-based data augmentation for conjoint studies at the analysis stage. \cite{ye2025lola} shows how LLMs enhance online learning algorithms for optimizing user engagement. We contribute to this emerging literature by demonstrating how LLMs can improve experimental design itself—enabling more precise treatment effect estimation and hypothesis testing, which is valuable for many social science experiments. 

\section*{Method}
\label{sec:methodology}
\addcontentsline{toc}{section}{Method}

To set the stage, consider a randomized experiment with $n$ units indexed by $i \in \{1, \ldots, n\}$, and potential outcomes $Y_i(1)$ under treatment and $Y_i(0)$ under
  control. The experimenters seek to estimate the average treatment effect $\tau = \E[Y_i(1) - Y_i(0)]$. Before the launch of the experiment, the experimenters observe a set
   of pre-treatment covariates $X_i$, and are interested in estimating the average treatment effect using the simple difference-in-means estimator $\widehat{\tau}_{simple} =
   \bar{Y}_1 - \bar{Y}_0$.
Although other estimators, such as regression with additional pre-treatment covariates, can improve
  the efficiency gain, it has more degrees of freedom and is considered less transparent and convincing than the difference-in-means estimator.\footnote{For example, if a
  study has a statistically insignificant treatment effect based on the simple difference-in-means estimator, even if the treatment effect becomes statistically significant
  after regression, the study may be less convincing compared to a carefully designed experimental study where the treatment effects are statistically significant based on
  the simple difference-in-means estimator.} The experimenters are evaluating, given the set of $X_i$ that they observe for each unit at the design stage, whether they
  should use stratification, and if so, how to conduct it effectively to improve experimental design and reduce the variance.

\subsection*{Optimal Stratification}
\addcontentsline{toc}{subsection}{Optimal Stratification}

Our proposed stratification method leverages insights from \cite{bai2022optimality}, who establishes that the optimal single-dimensional score for stratification is:
\begin{equation}
g^*(X) = \E[Y_i(1) + Y_i(0) | X_i = X]
\label{eq:optimal_index}
\end{equation}
A matched-pair design based on $g^*(X)$ minimizes the variance of the treatment effect estimator among all possible stratification schemes. Intuitively, units with similar values of $g^*(X)$ have similar expected outcomes under both treatment conditions, making them ideal candidates for comparison. The fundamental challenge is that $g^*(X)$ is not observed ex-ante, and therefore researchers must approximate this score using observed pretreatment covariates.

Faced with this challenge, experimenters typically resort to various approaches, each with important limitations. The most common experiment design, as indicated by the AEA RCT Registry, still remains to be simple randomization. Simple randomization, while unbiased and straightforward to implement, forgoes potential efficiency gains from incorporating covariate information at the design stage.\footnote{Both pre-experiment stratification and post-experiment regression adjustment can improve precision, but with different trade-offs. Stratification provides balance guarantees in finite samples and transparent pre-specification, while regression adjustment can achieve greater flexibility and handle continuous covariates more naturally. However, regression adjustment's validity depends on correct model specification for inference (though not for unbiasedness), and many fields still prefer design-based inference from stratified experiments for its robustness and interpretability.} When researchers decide to run stratified experiments, very often they rely on one or two covariates, typically chosen based on their prior beliefs or some common demographic variables such as age and gender.\footnote{All experiments analyzed by \cite{bai2022optimality} either use simple randomization, or stratify on at most two variables.} This approach can yield substantial efficiency gains when the variables used are highly predictive (such as baseline variables) of the outcome, but leaves additional efficiency gains on the table when other covariates also provide additional information to predict the outcome. One reason that researchers typically choose to stratify based on one or two variables is that matching on all variables quickly becomes infeasible as the number of covariates grows, creating strata with insufficient units for within-stratum randomization.\footnote{Re-randomization methods that reject imbalanced allocations can achieve covariate balance across multiple dimensions, but require specifying balance criteria and may complicate inference if not properly accounted for in the analysis.}

More sophisticated methods bring their own limitations. Mahalanobis distance matching \citep{greevy2004optimal}, a common choice for matched-pair designs, pairs units to minimize  

\[
d_M(X_i, X_j)=\sqrt{(X_i-X_j)^\prime \Sigma^{-1}(X_i-X_j)},
\]
where \(\Sigma\) is the covariance matrix. Although this procedure balances all covariates simultaneously, it weights them by their variance rather than by their predictive relevance, potentially over-emphasizing unimportant dimensions and underweighting crucial predictors. An alternative is to run a pilot study, estimate \(g^{*}(X)\) empirically, and use these estimates to create strata for the main experiment. While statistically principled, this strategy demands substantial time and resources, making it infeasible for many applications. Moreover, if the pilot sample is small, it may be too uninformative to define useful strata \citep{bai2022optimality}.

Given that each of these standard approaches faces limitations when confronted with experimental settings featuring rich covariate spaces, the gap between theoretical optimality and practical feasibility has sometimes left researchers without clear guidance on how to leverage the increasingly rich pre-treatment data available in many experimental contexts. This methodological gap motivates our proposed method: using Large Language Models to approximate the optimal prognostic score directly.

\subsection*{Generative Stratification}
\addcontentsline{toc}{subsection}{Generative Stratification}

We propose leveraging Large Language Models to approximate the optimal prognostic score $g^*(X)$ directly. This approach exploits the capacity of LLMs, trained on extensive corpora of human-generated text, to internalize complex representations of relationships between individual characteristics and behavioral outcomes across diverse contexts. 

Our approach rests on the observation that even if LLM prediction is imperfect, if it is correlated with the true optimal prognostic score, it can yield substantial efficiency gains. Consider a proxy $\widehat{g}(X)$ for the optimal prognostic score $g^*(X) = \mathbb{E}[Y(1) + Y(0)|X]$. For matched-pair designs, the efficiency gain from stratification depends on how well our proxy captures the conditional expectation of potential outcomes. Following \citet{bai2022optimality} and under mild assumptions\footnote{(i) 1:1 matched-pair randomization with ignorability, (ii) finite second moments of $Y(d)$, and (iii) pairing by an admissible index $h(X)$ with non-degenerate within-pair variance, Lipschitz conditional means in $h(X)$, and $\E[h(X)^2]<\infty$.} when units are paired and then randomized based on a prognostic score $h(X)$, the asymptotic distribution of $\sqrt{n}\,\big(\widehat{\tau}_{\text{paired} }- \tau)\xrightarrow{d} N (0, \mathbb{V}_{paired}^h)$, with 
\begin{equation}
\mathbb{V}_{paired}^h = \text{Var}[Y(1)] + \text{Var}[Y(0)] - \frac{1}{2}\text{Var}(\mathbb{E}[Y(1)+Y(0)|h(X)])
\end{equation}
In contrast, under standard simple randomization, the asymptotic distribution of $\sqrt{n}\,\big(\widehat{\tau}_{\text{simple} }- \tau)\xrightarrow{d} N (0, \mathbb{V}_{simple})$, with $\mathbb{V}_{simple} = \frac{1}{2}\left(\text{Var}[Y(1)] + \text{Var}[Y(0)]\right)$. Therefore, pairing reduces variance by capturing between-pair heterogeneity in the sum of potential outcomes:
\begin{equation}\label{eqn:efficiency_gain}
\frac{\mathbb{V}_{paired}^h}{\mathbb{V}_{simple}} = 1 - \frac{\text{Var}(\mathbb{E}[Y(1)+Y(0)|\widehat{g}(X)])}{2(\text{Var}[Y(1)] + \text{Var}[Y(0)])}
\end{equation}

The efficiency gain depends directly on the proportion of total variance in $Y(1) + Y(0)$ explained by our proxy $\widehat{g}(X)$ through its conditional expectation—an $R^2$ measure of outcome variance captured by proxy-based stratification. Even imperfect proxies can yield meaningful efficiency improvements when the correlation $\rho = \text{Corr}(\widehat{g}(X), g^*(X))$ is moderately high, as this typically translates to substantial explained variance.

The procedure incorporates inherent safeguards through its operation at the design stage. First, since the efficiency gain depends on explained variance rather than unbiased prediction, systematic over- or under-prediction does not compromise stratification quality as long as it explains the between-strata variation.
In the worst case where $\widehat{g}_{LLM}$\footnote{Note that negative correlation can also be helpful because sorting by a negatively correlated score still groups units with similar
characteristics together, thereby explaining between-strata variation.} constitutes pure noise uncorrelated with true potential outcomes, the conditional variance term vanishes and we effectively recover simple randomization with $\mathbb{V}_{paired}/\mathbb{V}_{simple} \approx 1$. The method thus cannot perform worse than simple randomization in expectation ex-ante \citep{athey2017econometrics}\footnote{Although it is plausible for it to perform worse ex-post.}. Most importantly, within-pair randomization guarantees preservation of unbiasedness, $\mathbb{E}[\widehat{\tau}_{paired}] = \tau$, regardless of the quality of $\widehat{g}_{LLM}$. Prediction errors affect the magnitude of efficiency gain but never compromise the validity of the experiment.

\subsection*{Implementation Procedure}
\addcontentsline{toc}{subsection}{Implementation Procedure}

We implement Generative Stratification in five streamlined steps that convert rich covariate information into an actionable stratification scheme.

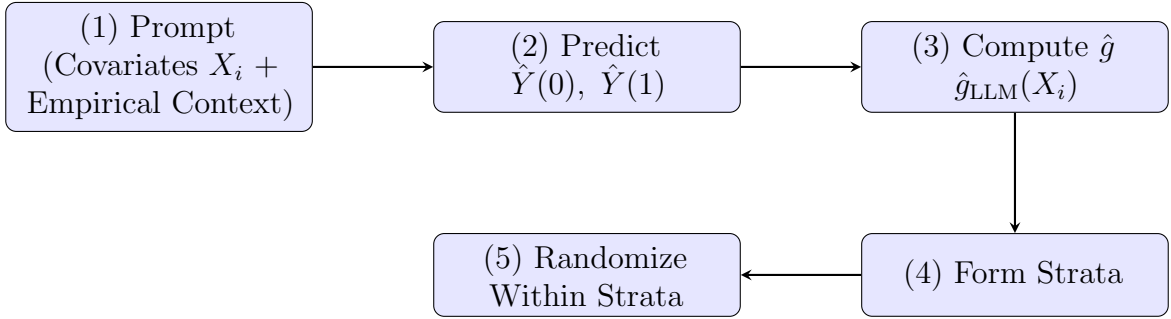
\begin{figure}[H]
    \centering
    \begin{tikzpicture}[
        node distance=1.6cm,
        process/.style={rectangle, rounded corners, draw=black, fill=blue!10, text width=3.8cm, align=center, minimum height=1.1cm},
        arrow/.style={->, >=stealth, thick}
    ]
        \node (context) [process] {(1) Prompt \\ (Covariates $X_i$ + Empirical Context)};
        \node (predict) [process, right=of context] {(2) Predict \\ $\widehat{Y}(0),~\widehat{Y}(1)$};
        \node (index) [process, right=of predict] {(3) Compute $\widehat{g}$ \\ $\widehat{g}_{\text{LLM}}(X_i)$};
        \node (strata) [process, below=of index] {(4) Form Strata};
        \node (randomize) [process, left=of strata] {(5) Randomize \\ Within Strata};
        \draw [arrow] (context) -- (predict);
        \draw [arrow] (predict) -- (index);
        \draw [arrow] (index) -- (strata);
        \draw [arrow] (strata) -- (randomize);
    \end{tikzpicture}
    \caption{Graphical overview of the five‐step Generative Stratification procedure.}
    \label{fig:gs_flow}
\end{figure}

\textbf{Step 1: Provide Context for LLM Simulation} For each experimental unit $i$ with covariates $X_i$, we construct a structured natural language description of the individual characteristics and experimental context. This description systematically incorporates demographic information, background characteristics, behavioral indicators, and contextual factors relevant to the outcome of interest. The experimental scenario, including detailed descriptions of control and treatment conditions, provides essential context for the LLM to generate informed predictions. This structured approach ensures consistent information extraction while maintaining flexibility to accommodate diverse covariate types and experimental settings. The appendix describes the templates we use in the study. For this exercise, we set the model to be gpt-4.1-2025-05-14 and the temperature to be $0$ to reduce variability conditional on covariates.

\textbf{Step 2: Outcome Prediction.} The LLM generates predictions for potential outcomes under both experimental conditions. For each unit, the model predicts $\widehat{Y}_i(0)$ under the control condition and $\widehat{Y}_i(1)$ under the treatment condition, leveraging its learned representations to synthesize the multidimensional covariate information. 

\textbf{Step 3: Score Construction.} Following the theoretical optimum established by \citet{bai2022optimality}, for standard experiment with treatment probability $p=0.5$, we compute the prognostic score as:
\begin{equation}
\widehat{g}_{LLM}(X_i) = \widehat{Y}_i(1) + \widehat{Y}_i(0)
\end{equation}
For designs with unequal treatment assignment probabilities $p \neq 0.5$, we employ the probability-weighted variant:
\begin{equation}
\widehat{g}_{LLM}(X_i) = \frac{\widehat{Y}_i(1)}{p} + \frac{\widehat{Y}_i(0)}{1-p}
\end{equation}

\textbf{Step 4: Stratum Formation.} This estimated score can be used in several different ways, depending on researchers' prior domain knowledge and experience with stratification. If researchers had been reluctant to use stratification, and this reluctance is partly driven by the complexity of deciding which variables to stratify on, we recommend simply sorting this score and matching units into pairs or sets of four.

If researchers intend to use other stratification methods because they are uncertain about the quality of this LLM-simulated score, they can consider treating this score as an additional covariate to incorporate into their intended stratification methods. If this score can capture complex nonlinear interactions among the existing covariates, or capture additional high-dimensional covariates that the researchers were not able to include, then it can further improve experimental efficiency. 

\textbf{Step 5: Randomization.} Within each stratum, we randomly assign units to treatment according to the target allocation rate.

\section*{Empirical Application}
\label{sec:empirical}
\addcontentsline{toc}{section}{Empirical Application}

\subsection*{Datasets}
\addcontentsline{toc}{subsection}{Datasets}

Our method is going to be most useful in settings where the researchers have access to a set of variables and it is not obvious which covariates to stratify on. To demonstrate the value of our method, we considered a set of 10 experiments that were systematically selected by \cite{bai2022optimality} to test different stratification methods. Among those 10 experiments, we were able to find datasets for 7 experiments. We removed one experiment that only has one covariate, which is obvious to stratify on, and two experiments with panel data, which one can stratify based on lagged outcomes. The remaining four experiment datasets are \cite{abel2020value}, \cite{barrera2019cct}, \cite{gerber2020million}, and \cite{demel2019labor}. 

\subsection*{Evaluation Design}
\addcontentsline{toc}{subsection}{Evaluation Design}

To evaluate our method, we conduct a simulation study that largely follows \cite{bai2022optimality} that evaluates the performance of the method. To make the evaluation exercise computationally feasible, we first sample $1,000$ observations from each study and then conduct simulation exercise based on these samples.\footnote{With the exception of \cite{demel2019labor}, which has fewer than $1,000$ observations and we selected $400$ observations.} The sample of each study can be represented as $( Y_i^{*}, D_i^{*}, X_i^{*})$, where $Y_i^{*}$ denotes the observed outcome, $D_i^{*}$ denotes the treatment status, and $X_i^{*}$ denotes the covariates used in the paper. Since only one of the potential outcomes is observed, to conduct the evaluation exercise we also need to impute the unobserved potential outcome so we have the full $(Y_i^*(1), Y_i^*(0), X_i^*)$. To make the imputation realistic, we estimate the heterogeneous treatment effect $Y_i^*(1) - Y_i^*(0)$ given $X_i^*$, and then impute the missing outcome by combining it with the observed outcome, and treat it as the ground truth to evaluate different experimental designs.\footnote{This imputation exercise is necessary only for the evaluation process and is not needed for implementing stratification.} Importantly, these imputed values are used only for evaluation purposes and are not needed for implementing Generative Stratification. They were not used by the LLM during stratification, as the LLM makes predictions based solely on pre-treatment covariates and treatment descriptions, ensuring our evaluation tests its ability to identify predictive patterns from pre-treatment information alone. For each study with $N$ observations, we conduct 3000 replications, where for each replication, we simulate new data by drawing $N$ units from the empirical distribution with replacement. Then we summarize the distribution of the ATE estimate across these replications under different stratification or estimation methods. 

\subsection*{Generative stratification versus simple randomization}
\addcontentsline{toc}{subsection}{Generative stratification versus simple randomization}

We first show the improvement when researchers use our stratification method to sort units into matched pair versus simple randomization in MSE. We also compare our method against the stratification method that was actually used by researchers in these published papers. For example, \cite{abel2020value} stratified by gender, \cite{barrera2019cct}  stratified by both grade and gender, \cite{gerber2020million}  stratified by state, 
and \cite{demel2019labor}  stratified by region and sector. A common theme in these cases is that the researchers only selected one or two variables, highlighting the empirical challenge of using all available covariates for stratification. 

Table \ref{tab:mse_llm_vs_original} reports the MSE of the difference-in-mean estimator under (1) simple randomization, (2) the stratification method used in the original paper, and (3) our generative stratification method. In all studies, generative stratification outperforms simple randomization by $10\%-50\%$. In 3 out of 4 studies, generative stratification substantially improves the performance relative to the original method. In \cite{barrera2019cct} where the original stratification uses grade that is highly predictive of the education outcome, our study does not improve the results but the difference is insignificant and small, showing that our method is on average beneficial and has limited risk.

\providecommand{\msecellsep}{0.25ex}
\providecommand{\msecell}[2]{%
  \begingroup
  \renewcommand{\arraystretch}{1.0}%
  \begin{tabular}[t]{@{}c@{}}%
    #1\\[-\msecellsep]{\small[#2]}%
  \end{tabular}%
  \endgroup
}

\begin{table}[H]
\centering
\caption{\textbf{MSE Improvement Achieved by LLM Stratification}}
\label{tab:mse_llm_vs_original}
{\renewcommand{\arraystretch}{1.20}%
\begin{adjustbox}{max width=\textwidth}
{\large
\begin{tabular}{@{}lccccc@{}}
\toprule
& \multicolumn{3}{c}{\textbf{MSE}} & \multicolumn{2}{c}{\textbf{Relative Improvement}} \\
\cmidrule(lr){2-4} \cmidrule(lr){5-6}
\textbf{Study}
& \textbf{Simple}
& \textbf{Original}
& \textbf{Generative}
& \textbf{Simple (\%)}
& \textbf{Original (\%)} \\
& (1) & (2) & (3) & (4) & (5) \\
\midrule
Abel (2020)   & \msecell{0.19422}{0.18433, 0.20411} & \msecell{0.19118}{0.18179, 0.20057} & \msecell{0.12759}{0.12103, 0.13415} & 34.3 & 33.3 \\
\addlinespace[2pt]
Barrera (2019)   & \msecell{0.00071}{0.00068, 0.00075} & \msecell{0.00063}{0.00060, 0.00066} & \msecell{0.00063}{0.00060, 0.00066} & 11.2 & -0.5 \\
\addlinespace[2pt]
Gerber (2020)   & \msecell{0.73688}{0.70130, 0.77246} & \msecell{0.68092}{0.64639, 0.71544} & \msecell{0.35656}{0.33768, 0.37544} & 51.6 & 47.6 \\
\addlinespace[2pt]
Mel (2019)   & \msecell{0.00844}{0.00801, 0.00886} & \msecell{0.00831}{0.00789, 0.00872} & \msecell{0.00719}{0.00682, 0.00756} & 14.8 & 13.5 \\
\bottomrule
\end{tabular}
}
\end{adjustbox}
}
\\[3pt]
\parbox{\textwidth}{\footnotesize\emph{Notes.} Columns (1)–(3) report MSE with 95\% bootstrap confidence intervals in brackets.
Column (4) is computed as $(\text{(1)}-\text{(3)})/\text{(1)}\times 100$, 
and Column (5) is computed as $(\text{(2)}-\text{(3)})/\text{(2)}\times 100$.}
\end{table}

  Table \ref{tab:se} demonstrates the efficiency gains in standard errors. These improvements mirror the MSE reductions, with standard errors under Generative Stratification being on average 15.3\% lower than under simple randomization and 12.6\% lower than under the original stratification methods.\footnote{Standard errors for the generative stratification were calculated using the adjusted variance estimator following \cite{bai2022optimality}: $\widehat{\varsigma}^2_{h,n}
\;=\;
\hat{\sigma}_n^2(1)\;+\;\hat{\sigma}_n^2(0)\;-\;\tfrac{1}{2}\,\hat{\rho}_n\;+\;\tfrac{1}{2}\big(\hat{\mu}_n(1)+\hat{\mu}_n(0)\big)^2
$. } The reduced standard errors translate directly to improved statistical power, allowing researchers to detect smaller treatment effects with the same sample size or achieve equivalent power with smaller samples.

\begin{table}[H]
\centering
\caption{\textbf{SE Improvement Achieved by LLM Stratification}}
\label{tab:se}
{\renewcommand{\arraystretch}{1.20}%
\begin{adjustbox}{max width=\textwidth}
{\large
\begin{tabular}{@{}lccccc@{}}
\toprule
& \multicolumn{3}{c}{\textbf{SE}} & \multicolumn{2}{c}{\textbf{Relative Improvement}} \\
\cmidrule(lr){2-4} \cmidrule(lr){5-6}
\textbf{Study}
& \textbf{Simple}
& \textbf{Original}
& \textbf{Generative}
& \textbf{Simple (\%)}
& \textbf{Original (\%)} \\
& (1) & (2) & (3) & (4) & (5) \\
\midrule
Abel (2020)   & 0.43184 & 0.43103 & 0.35796 & 17.1 & 17.0 \\
\addlinespace[2pt]
Barrera (2019)   & 0.02674 & 0.02501 & 0.02525 & 5.6 & -1.0 \\
\addlinespace[2pt]
Gerber (2020)   & 0.87230 & 0.83118 & 0.60684 & 30.4 & 27.0 \\
\addlinespace[2pt]
Mel (2019)   & 0.09132 & 0.09091 & 0.08394 & 8.1 & 7.7 \\
\midrule
\textbf{Average}   &  &  & & \textbf{15.3} & \textbf{12.6} \\
\bottomrule
\end{tabular}
}
\end{adjustbox}
}
\\[3pt]
\parbox{\textwidth}{\footnotesize\emph{Notes.} Columns (1)–(3) report the mean of analytic standard error. Column (4) is computed as $(\text{(1)}-\text{(3)})/\text{(1)}\times 100$, 
and Column (5) is computed as $(\text{(2)}-\text{(3)})/\text{(2)}\times 100$.}
\end{table}

\subsection*{Comparison with other alternatives}
\addcontentsline{toc}{subsection}{Comparison with other alternatives}

The aforementioned method has only considered using the generative score for ordering units. To further improve the efficiency gain, the score  can also be combined with existing stratification rules, hence further outperforming other post-experiment adjustment alternatives. Using this prognostic score as an additional covariate can be valuable because it can capture interactions between existing covariates that would otherwise be difficult to model or unstructured data that are difficult to incorporate at the design stage. Additionally, since different covariates contribute differently to the final outcome, this also represents a way to reweight the existing covariates, giving higher weights to covariates that may be more predictive of the final outcome. For example, one easy way to combine them is calculate weighted average of difference distances, which trades off closeness in $\widehat g$ and Mahalanobis balance on $X$, creating strata to minimize the pair distance 
\begin{equation}
c_\lambda(i,j)
=
\lambda\,\frac{\big(\widehat g_i-\widehat g_j\big)^2}{\widehat s_g^2}
\;+\;
(1-\lambda)\,\big(X_i-X_j\big)^{\!\top}\widehat\Sigma_X^{-1}\big(X_i-X_j\big),
\qquad \lambda\in[0,1],
\label{eq:pair_cost}
\end{equation}
where $\widehat s_g^2$ is the sample variance of $\widehat g$ (to standardize scales) and $\widehat\Sigma_X$ is the sample covariance of $X$. Setting $\lambda=1$ yields pure $\widehat g$–based pairing; $\lambda=0$ yields standard Mahalanobis pairing \citep{greevy2004optimal}. 

The choice of $\lambda$ reflects researchers' confidence in the accuracy of the simulation score. When LLM development is at a relatively early stage and researchers have no prior knowledge about its ability, they can start with a low value. As the LLM becomes better or gains access to better training data in the future, $\lambda$ can be further increased. For our empirical illustration, we adopt a simple heuristic of $\lambda = \frac{1}{k+1}$, where $k$ is the dimension of X used for the Mahalanobis calculation. This choice treats the LLM-generated score as roughly equivalent in weight to a single independent covariate to avoid over-reliance on the simulated score. In practice, researchers could adjust it based on domain expertise and experience about the relative informativeness of the LLM predictions versus observed covariates. 
Table \ref{tab:mse_llm_gain_2} reports the MSE under three methods: (1) regression, which performs simple randomization followed by OLS estimation with covariates, (2) matched pairs using Mahalanobis distance on all covariates $X$ (MP X), and (3) matched pairs using our hybrid distance that incorporates the LLM-generated score (MP LLM X). The hybrid approach consistently outperforms both alternatives, achieving approximately $50\%$ improvement over the regression method and $2\%-13\%$ improvement over standard Mahalanobis pairing. These results demonstrate the substantial value of pre-experiment design over post-experiment adjustment, and show that incorporating the LLM-generated prognostic score further enhances stratification performance even when all covariates are already being utilized.

\providecommand{\msecellsep}{0.25ex}
\providecommand{\msecell}[2]{%
  \begingroup
  \renewcommand{\arraystretch}{1.0}%
  \begin{tabular}[t]{@{}c@{}}%
    #1\\[-\msecellsep]{\small[#2]}%
  \end{tabular}%
  \endgroup
}

\begin{table}[H]
\centering
\caption{\textbf{MSE: Regression vs Matched-Pairs (Mahalanobis) vs Matched-Pairs (Hybrid with LLM Score)}}
\label{tab:mse_llm_gain_2}
{\renewcommand{\arraystretch}{1.20}%
\begin{adjustbox}{max width=\textwidth}
{\large
\begin{tabular}{@{}lcccccc@{}}
\toprule
& \multicolumn{3}{c}{\textbf{MSE}} & \multicolumn{2}{c}{\textbf{Relative Improvement}} \\
\cmidrule(lr){2-4} \cmidrule(lr){5-6}
\textbf{Study}
& \textbf{Regression}
& \textbf{MP X}
& \textbf{MP LLM X}
& \textbf{Regression (\%)}
& \textbf{MP X (\%)} \\
& (1) & (2) & (3) & (4) & (5) \\
\midrule
Abel (2020)
& \msecell{0.13312}{0.12664, 0.13959}
& \msecell{0.06722}{0.06157, 0.07287}
& \msecell{0.06307}{0.05740, 0.06875} 
& 52.6 & 6.2 \\
\addlinespace[2pt]
Barrera (2019)
& \msecell{0.00064}{0.00060, 0.00067}
& \msecell{0.00031}{0.00028, 0.00034}
& \msecell{0.00030}{0.00028, 0.00033} 
& 52.4 & 2.8 \\
\addlinespace[2pt]
Gerber (2020)
& \msecell{0.39498}{0.37517, 0.41479}
& \msecell{0.21148}{0.19879, 0.22418}
& \msecell{0.18477}{0.16905, 0.20050} 
& 53.2 & 12.6 \\
\addlinespace[2pt]
Mel (2019)
& \msecell{0.00694}{0.00658, 0.00730}
& \msecell{0.00369}{0.00335, 0.00403}
& \msecell{0.00344}{0.00312, 0.00375} 
& 50.5 & 7.0 \\
\bottomrule
\end{tabular}%
}%
\end{adjustbox}
}
\\[3pt]
\parbox{\textwidth}{\footnotesize\emph{Notes.} Columns (1)–(3) report MSE with 95\% bootstrap confidence intervals in brackets. Column (4) is $(\text{(1)}-\text{(3)})/\text{(1)}\times 100$, Column (5) is $(\text{(2)}-\text{(3)})/\text{(2)}\times 100$.}
\end{table} 

\section*{Conclusion}
\label{sec:conclusion}
\addcontentsline{toc}{section}{Conclusion}

This paper introduces \emph{Generative Stratification}: a simple, theory-guided way to compress rich, heterogeneous covariates into a one-dimensional prognostic score using Large Language Models, and then stratify by that score. It is easy to deploy (predict, sort, pair), accounts for various data types (numeric, categorical, text), and integrates naturally with matched-pair and other standard stratification designs. Across our empirical applications, replacing simple randomization with our LLM-based ordering consistently improves precision; combining the score with other standard stratification methods (e.g., the hybrid cost in Eq.~\eqref{eq:pair_cost}) can yield further gains.

While our empirical applications demonstrate meaningful improvements, these applications are constrained by the limited pretreatment data available in existing experimental datasets. The value of \emph{Generative Stratification} can potentially increase further when researchers have access to richer, unstructured pretreatment information—such as text responses, interview transcripts, or detailed behavioral histories—that traditional stratification methods cannot easily incorporate. Moreover, as LLM capabilities continue to advance in understanding and predicting individual-level heterogeneity, the efficiency gains from our approach will also improve. Thus, the method we introduce today has the potential to become increasingly valuable as both data availability and model capability grow.

The paper contributes to the topic of LLM simulation by presenting a \emph{responsible and low-risk} use case of LLMs in experimentation. First, the LLM affects only design, not identification: unbiasedness follows from within-stratum randomization. Second, efficiency depends on correlation with the optimal score, not on perfect calibration. We hope that as researchers continue to work on LLM models and improve their prediction ability, the value of this method can be further amplified to make experiments more efficient in the future.

\newpage
\bibliographystyle{apalike}
\singlespacing
\bibliography{references}

\appendix

\section*{Appendix: Simulation Template}\label{appendix:template}
\addcontentsline{toc}{section}{Appendix A: Simulation Template}

\begin{lstlisting}[language={},title={Simulation Template}]
You are an AI assistant tasked with predicting outcomes for individuals in an experiment based on their characteristics and treatment conditions.

(*\textcolor{sectioncolor}{<experiment\_background>}*)
[Detailed description of the experimental context, including:
 - Study setting and population
 - Treatment intervention details
 - Control condition description
 - Relevant background information]
(*\textcolor{sectioncolor}{</experiment\_background>}*)

(*\textcolor{sectioncolor}{<outcome\_definition>}*)
The outcome to predict is: [Specific outcome description]
(*\textcolor{sectioncolor}{</outcome\_definition>}*)

(*\textcolor{sectioncolor}{<variable\_descriptions>}*)
[List of available covariates with descriptions:
 - variable_name: Description of what this variable measures
(*\textcolor{sectioncolor}{</variable\_descriptions>}*)

(*\textcolor{sectioncolor}{<individual\_characteristics>}*)
[Specific values for this individual:
 - variable_name: (*\textcolor{placeholdercolor}{\{\{variable\_name\}\}}*)
(*\textcolor{sectioncolor}{</individual\_characteristics>}*)

(*\textcolor{sectioncolor}{<treatment\_condition>}*)
(*\textcolor{placeholdercolor}{\{\{treatment\_status\}\}}*)
(*\textcolor{instructioncolor}{[Filled with: "This individual RECEIVES the treatment" or}*)
(*\textcolor{instructioncolor}{"This individual DOES NOT receive the treatment (control group)"]}*)
(*\textcolor{sectioncolor}{</treatment\_condition>}*)

Based on the information provided, predict the outcome for this individual under both the control condition (does not receive treatment) and the treatment condition (receives treatment).

Your response should contain only two numbers:
1. The predicted [outcome_type] under the control condition
2. The predicted [outcome_type] under the treatment condition

Format your response EXACTLY as follows:
(*\textcolor{sectioncolor}{<prediction>}*)
[Control prediction]
[Treatment prediction]
(*\textcolor{sectioncolor}{</prediction>}*)

Example response:
(*\textcolor{sectioncolor}{<prediction>}*)
[example_value_control]
[example_value_treatment]
(*\textcolor{sectioncolor}{</prediction>}*)

Do not provide any explanation or commentary.
\end{lstlisting}

\end{document}